\documentclass{article}

\pdfoutput=1

\usepackage[preprint]{neurips_2020}

\usepackage[utf8]{inputenc} %
\usepackage[T1]{fontenc}    %
\usepackage{url}            %
\usepackage{hyperref}       %
\usepackage{booktabs} %

\usepackage{multirow}
\usepackage{amsfonts}       %
\usepackage{nicefrac}       %
\usepackage{microtype}      %
\usepackage{appendix}

\usepackage{amsmath}
\usepackage{amsthm}
\usepackage{amssymb}

\usepackage{ifthen}
\usepackage{xcolor}
\definecolor{darkred}{rgb}{.7,0,0}
\definecolor{darkgreen}{rgb}{0,.5,0}
\definecolor{darkblue}{rgb}{0,0,.8}
\definecolor{darkcyan}{rgb}{0,0.6,.6}
\definecolor{darkorange}{rgb}{.8,.4,0}
\definecolor{gray}{rgb}{.4,.4,.4}

\newcommand{\needcite}[1]{\textcolor{darkorange}{\emph{[citation needed\ifthenelse{\equal{#1}{}}{}{: #1}]}}}

\usepackage{graphicx}
\usepackage{pgf}
\usepackage{layouts}

\usepackage{etoolbox}
\usepackage{siunitx}
\usepackage{selectp}

\usepackage{algorithm}
\usepackage{subcaption}
\usepackage{listings} %

\usepackage{letltxmacro}
\newcommand*{\SavedLstInline}{}
\LetLtxMacro\SavedLstInline\lstinline
\DeclareRobustCommand*{\lstinline}{%
  \ifmmode
    \let\SavedBGroup\bgroup
    \def\bgroup{%
      \let\bgroup\SavedBGroup
      \hbox\bgroup
    }%
  \fi
  \SavedLstInline
}
\newcommand{\code}[1]{\lstinline[mathescape=true]{#1}}
\title{Learning To Prove From Synthetic Theorems}

\author{%
  Eser Aygün \\
  DeepMind \\
  \texttt{eser@google.com} \\
  \And Zafarali Ahmed \\
  DeepMind \\
  \texttt{zaf@google.com} \\
  \And Ankit Anand \\
  DeepMind \\
  \texttt{anandank@google.com} \\
  \AND Vlad Firoiu \\
  DeepMind \\
  \texttt{vladfi@google.com} \\
  \And Xavier Glorot \\
  DeepMind \\
  \texttt{glorotx@google.com} \\
  \And Laurent Orseau \\
  DeepMind \\
  \texttt{lorseau@google.com} \\
  \And Doina Precup \\
  DeepMind \\
  \texttt{doinap@google.com} \\
  \And Shibl Mourad \\
  DeepMind \\
  \texttt{shibl@google.com} \\
}

\begin{document}

\maketitle
\begin{abstract}
 A major challenge in applying machine learning to automated theorem proving is the scarcity of training data, which is a key ingredient in training successful deep learning models. To tackle this problem, we propose an approach that relies on training with synthetic theorems, generated from a set of axioms. We show that such theorems can be used to train an automated prover and that the learned prover transfers successfully to human-generated theorems. We demonstrate that a prover trained exclusively on synthetic theorems can solve a substantial fraction of problems in TPTP, a benchmark dataset that is used to compare state-of-the-art heuristic provers. Our approach outperforms a model trained on human-generated problems in most axiom sets, thereby showing the promise of using synthetic data for this task.   
\end{abstract}
\section{Introduction}

It has long been a dream of computer scientists to automate mathematics by building powerful automated theorem provers (ATPs) \citep{newell&herbert56,lenat77}. The successes of deep learning methods, particularly in computer vision \citep{he&al16}, natural language processing \citep{devlin&al19}, and in discrete games with perfect information \citep{silver&al16}, raise the possibility that similar methods could give rise to strong ATPs. An important reason for deep learning's success in these areas is the abundance of data in the form of large datasets of images \citep{russakovsky&al15}, text \citep{merity&al17}, or from simulated self-play in symmetric two-player games.

However, today's state-of-the-art ATPs, such as Vampire \citep{riazanov&voronkov02} and E \citep{schulz&al19}, rely mostly on finely-tuned heuristics discovered over the years and little on data. Many previous efforts that apply machine learning to theorem proving~\citep{loos&al17,jakubuv&urban17,crouse&al19} use human data for both training and testing and are focused on improving these existing heuristic provers rather than learning from scratch.

In this work, we explore an approach to ATP which is more akin to the data-powered deep learning methodology. We propose a process for training domain-specific provers {\em without using any theorems or proofs formulated by humans}. This process starts with an axiomatization of the domain at hand, and makes use of a synthetic theorem generator and a simple untrained prover to generate a large amount of proof data. This data is then used to train a better prover. 

Specifically, our synthetic theorem generator proceeds similarly to a theorem prover. It applies logical rules to the axioms of a given domain, and generates inferences that are implied by the axioms. Since the inferences are logically implied, they are always true given the axioms and can be used as theorems for learning a domain specific prover.

Our prover relies on a popular search algorithm, called \emph{saturation}. A key component of saturation is a cost function that evaluates the usefulness of available inferences during proof search. Similar to prior work \citep{loos&al17}, we formulate learning this cost function as a supervised task. However, unlike previous work, the data for this task is generated by a simple untrained prover on synthetic theorems rather than from human or ATP proofs of human-written theorems. Our empirical results show that the cost function learned through this methodology leads to saturation provers that can prove human-written theorems (from the popular Thousands of Problems for Theorem Proving (TPTP)~\citep{sutcliffe17})  without any exposure to human theorems or proof data. This result holds both for multi-layer perceptrons and graph neural networks. Additionally, the resulting provers compare favourably to versions which are both trained and tested on TPTP, further emphasizing the power of using large amounts of data, even if it comes from a very different, randomized distribution. Our provers also generate fewer clauses during inference, which holds promise for scaling to more complex mathematical theories.

\section{Background}
\label{sec:background}

\subsection{First-order logic}

First-order logic is a formal language that sits somewhere between propositional logic used by SAT solvers and higher-order logic used by mathematicians and interactive proof assistants. It allows one to construct \emph{formulas} that state facts about arbitrary objects such as numbers, sets or people using the universal ($\forall$) and existential ($\exists$) quantifiers, and to derive new facts using certain inference rules. %

While first-order logic is less expressive than higher-order logic,
certain domains of mathematics, such as geometry and algebra, can be formulated as first-order theories. First-order logic has also been used for hardware and software verification~\citep{claessen&al12}, and symbolic reasoning~\citep{wittocx&al08}. Most importantly, first-order logic is the preferred language for most of the fully automated theorem provers (ATPs)~\citep{schulz&al19, riazanov&voronkov02}, largely due to the simplicity of its rules of inference. 

All first-order formulas can be transformed into a special form called conjunctive normal form (CNF). A CNF formula is a conjunction (``and'' operator, $\wedge$) of clauses which are themselves disjunctions (``or'' operator, $\vee$) of literals. Each literal is a negated or non-negated atomic formula, such as $\neg p(X)$ or $q(f(X, c))$. Atomic formulas are recursive structures of atomic terms ($f(X, c)$ and $c$) and variables ($X$). Top-level function symbols in atomic formulas ($p$ and $q$) are called \emph{predicates} and top-level function symbols in atomic terms ($f$ and $c$) are called \emph{functors}.

One particular proof technique, the \emph{resolution calculus}, and the associated \emph{saturation algorithm}, are at the core of many successful first-order ATPs.

\subsection{Resolution calculus}

Resolution calculus is a set of inference rules that operate on a CNF formula. It allows one to generate new clauses that are logically implied by either one or two of the existing clauses. Selecting an inference adds the new clause to the set of existing clauses and allows yet more clauses to be derived. Typically, a proof starts with a set of axioms or hypotheses and a negation of the desired conclusion. The end goal is to arrive at the empty clause, which is logically equivalent to false; doing so amounts to a proof-by-contradiction.

Resolution calculus is \emph{refutation-complete}~\citep{robinson65}, meaning that, if the given conjecture is in fact true, then there is always a sequence of inferences which derives the empty clause. The job of a resolution-based theorem prover is to pick the right inferences at each step. This can be very difficult, as the number of available inferences can grow quadratically with the number of clauses. For a formal treatment of these topics, we refer the reader to \cite{fitting12}.

\subsection{Saturation}

The saturation algorithm answers the question of which inference to pick at each step of the resolution calculus. It uses a \emph{cost function} to assign a cost to each of the available inferences independently. At each step, an inference with minimum cost is selected. Using a good cost function is crucial for efficiently searching the space of clauses. State-of-the-art provers like Vampire~\citep{riazanov&voronkov02} and E~\citep{schulz&al19} come equipped with multiple cost functions carefully crafted for certain types of theorems, and pick the right one either by trial-and-error or based on an initial analysis of the problem at hand.
Heuristic cost functions tend to be very cheap to evaluate, allowing saturation provers to rapidly churn through many clauses during proof search. 
In addition to the cost functions, there are several other components, such as the elimination of redundant clauses, that contribute to the success of saturation. Further details are provided in the Appendix~\ref{appendix:saturation}.

\section{Approach}
\label{sec:methodology}

In this work, we build a saturation prover that uses a deep neural network cost function and minimal heuristics.
As in previous work~\citep{loos&al17}, we train the network to predict the probability of whether a clause will appear in the proof. We obtain the training data by running our saturation prover without any machine learning on synthetically generated theorems. After each successful proof, we record the clauses generated during saturation and whether each clause was used or not in the proof. We train both multi-layer perceptron (MLP) and graph neural network (GNN) models. Once trained, the network is plugged back into the saturation prover as a cost function.

\subsection{Basic saturation prover}

We have implemented a bare-bones saturation prover that uses a small number of standard, domain-agnostic heuristics. This prover can use either a deep network cost function, or a default heuristic one (used to generate training data).

As the default cost function, we use a simple metric called \emph{clause weight}, which measures the size of a given clause. More precisely, every clause can be represented as a tree where the root node represents the clause, its children represent the literals and anything below represent atomic formulas, atomic terms and variables.
The clause weight is defined as the number of nodes in its tree representation.  For example, the weight of the clause $\neg p(X) \lor q(f(X, c))$ is 9, that is: one clause node, two literal nodes, two atomic formula nodes, two atomic term nodes and two variable nodes. Note that this is a very uninformative heuristic but trivial to compute.

The prover also implements a simple clause redundancy check, namely \emph{$\theta$-subsumption}, which is widely used in first-order ATPs. The pseudocode of our prover and further details  are given in Appendix~\ref{appendix:saturation}.
We have opted for building our own prover from scratch in order to measure the efficacy of our methodology in a ``sterile'' environment, untainted by the biases of complex heuristics. In practice, our approach could just as easily work with any saturation-based prover.

\subsection{Synthetic theorem generator} \label{sec:generator}

We obtain our training theorems using a simple but novel synthetic theorem generator, which we call the \emph{forward proposer} (FwdP). It starts from the clauses corresponding to the axioms of a specific domain of mathematics (e.g., geometry) and uses resolution calculus to infer new clauses. The choice among available resolutions is taken uniformly at random. The forward proposer keeps generating new clauses in this fashion for a certain number of steps, and uses the final clause it obtained in order to build a conjecture of the form: Axioms $\rightarrow$ Clause. As the process starts with the axioms and the clauses are generated by logical inferences, this conjecture is guaranteed to be a valid theorem in the given domain.

In order to force the proposer towards clauses with deeper proof trees, at each step (except the first) we only allow inferences that involve the last generated clause. This is known as \emph{linear resolution}, and despite this restriction it is known to be as powerful as full resolution~\citep{fitting12}.

It should also be noted that, although the resolution calculus is implication-complete~\citep{lee67} (\emph{i.e} the forward proposer is capable of generating (almost) any clause implied by the axioms)~\footnote{"Lee's theorem" states that for any clause $C$ that is implied by the axioms, there is a clause $C'$ that can be generated by the resolution calculus such that $C'$ \emph{subsumes} $C$. For more details, see \cite{fitting12}.}, there is a class of theorems that it cannot generate: those whose conclusion cannot be expressed as a single clause.

\subsection{From synthetic theorems to supervised learning}

We used axiom sets available in the TPTP library to generate theorems. These axioms come separated according to the different domains of mathematics and reasoning. We formed axiom sets by grouping axiom files that occur together in the same theorems. We then filtered out any axiom sets with more than 1000 axiom clauses or less than ten associated theorems. We also excluded any domains that required first-order equality (an additional feature that our prover does not implement currently). This left us with ten axiom sets, covering  field theory (FLD), geometry (GEO), number theory (NUM), group theory (GRP), set theory (SET) and knowledge representation (KRS).  Table~\ref{table:datasets} contains the details regarding the numbers of axioms and TPTP theorems in each dataset. 

For each axiom set, we generated 25,000 training theorems and 1000 validation theorems by running the forward proposer for ten steps on the axioms of the set. We then ran our basic saturation prover on both sets of problems. For every theorem proved, we recorded the clauses generated during the proof process, as well as the initial set of clauses. We divided these clauses into two sets: a positive example set, composed of the clauses that appeared in the proof, and a negative example set, composed of the clauses that did not appear in the proof. Finally, we sub-sampled the (much larger) negative set uniformly to get an equal number of positive and negative examples.

\begin{table}
    \centering
    \caption{\label{table:datasets}Axiom sets and human-generated theorems extracted from TPTP.}
    \begin{tabular}{
        c
        c
        *2{S[table-format=3.0]}
    }
    \toprule
    {Axiom Set} &                  {Domain} &  {Axioms} &  {Theorems} \\
    \midrule
           FLD1 &              Field Theory &        27 &          78 \\
           FLD2 &              Field Theory &        26 &         105 \\
           GEO6 &                  Geometry &        46 &         128 \\
           GEO7 &                  Geometry &        58 &          38 \\
           GEO8 &                  Geometry &        35 &         128 \\
           GEO9 &                  Geometry &        66 &          37 \\
           GRP5 &              Group Theory &         7 &          10 \\
           KRS1 &  Knowledge Representation &       108 &          41 \\
           NUM9 &             Number Theory &        42 &          30 \\
           SET1 &                Set Theory &        24 &          11 \\
    \bottomrule
    \end{tabular}
\end{table}

We used the training examples generated in the previous step to train models for predicting whether a clause will appear in the proof. Model inputs consisted of two parts: (i) the clause to be evaluated and (ii) the initial set of clauses. As the goal of the prover is to reach an empty clause, the initial set of clauses can be viewed as a ``goal'' that should be matched by the current clause.

\subsection{Representation \& model architecture}
\label{sec:representation}
We experimented with two neural network architectures: a multi-layer perceptron (MLP) with a flat set of features and a graph neural network (GNN) with a graphical representation of the inputs. 

\paragraph{MLP.} We used seven scalar features to represent each clause: number of negated literals, number of non-negated literals, number of atomic terms, number of distinct predicates, number of distinct functors, number of distinct variables and total number of variables. The features of the initial clauses were further aggregated via four different aggregation functions: sum, average, maximum and minimum. Finally, three more scalars were concatenated to these: the number of the step at which the clause to be evaluated was generated, the number of premises used in the inference that generated the clause (between 0 and 2 depending on the type of inference used) and the number of initial clauses. In total, there were 38 elements in the input vector. For every model and dataset pair, we tried learning rates of \SI{1e-2}, \SI{3e-3}, \SI{1e-3}, \SI{3e-4} and \SI{1e-4}, and five different random number generator seeds. The batch size was fixed to 4096. We opted for a five layer MLP with layer sizes 256, 64, 16, 4 and 1, and ReLU activation \citep{nair2010rectified}. We did not do a thorough architecture search to optimize these numbers. 

\paragraph{GNN.} We used a graphical representation of first-order formulas described in \citet{glorotlearning}. Different types of nodes were instantiated to represent entries of the CNF: a formula node, clause nodes, negated and non-negated literal nodes, term nodes, predicate nodes, functor nodes and variable nodes. The graph formed a tree except for the last three node type instances which can be shared across the different clauses or literals subtrees. An additional node type was created to identify the argument ordering of predicates (or functors): all literal (or term) nodes that were placed at the same argument index for a given predicate (or functor) were connected to the same index node. This representation was therefore invariant with respect to clause/literal/variable ordering; predicate/functor/variable naming and predicate/functor arguments ordering.
A last node type was created to represent the clause to evaluate for the saturation algorithm. A different (learnable) embedding vector was initialized for all node and edge types. They are shared as features for the nodes/edges instances. In addition, we used a global graph feature vector initialized to the input vector of the MLP from above.

We derived the implementation of our GNNs from interaction networks~\citep{battaglia2016interaction}. Graph features were updated by a series of message passing steps. Before each step, we concatenated the initial graph with the current one, forming a skip connection. The node, edge and global update functions were single-layer MLPs with a ReLU non-linearity followed by layer norm \citep{ba2016layer}. At each step we applied three procedures. First, the edge update: for each edge, its features, the node features of its two end points were concatenated and fed to the edge update function MLP. The output was the new edge feature vector. Then, the node update: for each node we concatenated its feature vector with the sum of the feature vectors of its incoming edges. The result was given as input to the node update function MLP and the output was used as the node features update. Lastly, global features update: this was done by concatenating to the current global features, the sum of the node feature vectors over the entire graph, as well as the sum of the edge feature vectors. The result was fed to the global update function. The output was the new global feature vector.
To compute the final output of the model, the initial global features (i.e. the MLP flat feature inputs) were concatenated with the final global features and then passed through a two-layer neural network with 64, 32 hidden units and ReLU activations. GNNs can easily overfit the data and can be slow to run. Therefore, we restricted the size of the update function MLPs layer to 16 units and used L2 regularization. We did a hyper-parameter search over learning rates and the L2 regularization coefficient. We used 5 message passing steps based on an initial evaluation on a subset of GEO8 theorems.

A cross-entropy loss for the binary classification task was applied to the outputs and we trained the networks using the Adam optimizer~\citep{kingma2014adam}.
We ran the training loops for a maximum of 10,000 epochs and used the validation accuracy to stop the training early, in cases where overfitting was observed.
We selected the models with the highest overall validation accuracies to be used in the evaluation step.

\subsection{Evaluation}

From the trained models, we built learned cost functions for the saturation prover, according to the formula $(1 - p) + w / M$, where $p$ is the probability of the clause appearing in the proof as predicted by the model, $w$ is the clause weight and $M$ is the scale hyper-parameter. We observed that using only the model prediction was not enough in general, as the models were not trained to learn the order in which the clauses must be generated. Mixing in a small amount of the clause weight kept the branching factor lower in the initial steps. Our approach can be viewed as making the cost function account not only for whether or not a clause would appear in a proof, but also for the computational burden introduced by the clause, in terms of space and time. For the experiments reported below, we chose $M=16$ by doing a hyper-parameter search on a subset of GEO8. Training models that eliminate $w$ completely from the cost function of the saturation prover is left for future work.

To compare our approach against one that trains from human-generated theorems, we also built datasets of clauses from the TPTP theorems themselves, following the same protocol. The source of these clauses was naturally limited to those human-generated theorems that can be proven by the basic saturation algorithm. As the number of examples was already limited, we did not attempt to create a separate validation set for this approach.

We evaluated all cost functions on the human-generated theorems from TPTP. We limited every proof attempt to 300 seconds. If the time limit was exceeded, we marked the attempt as a failure. For the successful attempts, we measured the time it took until completion and the total number of clauses generated in the process.

To get a baseline with which to compare all learned cost functions, we also ran E version 1.9 under the same conditions. E is a state-of-art open-source prover which has many options that enable and disable different features; we tested two different configurations: with no flags (E) and with the flag ``--auto'' (E --auto). The --auto flag enables many of the complex heuristics embedded in E. Therefore E --auto can be considered as representative of the state-of-the-art in first-order theorem proving.

\section{Empirical Results}
\label{sec:results}

The main goal of our experiments was to measure the knowledge transfer from synthetic theorems (FwdP) to human-generated theorems (TPTP) and compare this to the transfer within human-generated theorems. Here, we refer to the models trained on synthetic theorems as MLP-on-FwdP and GNN-on-FwdP, and models trained on human-generated theorems as MLP-on-TPTP and GNN-on-TPTP.

\paragraph{Successful transfer from FwdP to TPTP.} All of the trained models were able to improve upon the basic saturation prover that was used to generate the classification examples (Table~\ref{table:success-counts}). In particular, MLPs solved 71 more problems than than the basic prover while GNNs solved 41 more problems. Even though GNNs had higher accuracies than MLPs for 9 out of 10 datasets (Table~\ref{table:accuracies-side-by-side}), they solved 30 fewer problems. Note also that the learned cost functions obtained better performance than the baseline E in 8 out of 10 domains, even though they solved slightly fewer problems overall.
\begin{table}
    \centering
    \caption{\label{table:accuracies-side-by-side}Accuracy, precision and recall on the validation set for the MLP and GNN models that are trained on synthetic theorems. As expected, the more expressive GNNs show better performance.}
    
    \begin{tabular}{
        c
        *6{S[table-format=1.3,round-mode=places,round-precision=3]}
    }
    \toprule
    &  \multicolumn{3}{c}{MLP-on-FwdP} & \multicolumn{3}{c}{GNN-on-FwdP} \\
    \cmidrule(lr){2-4} \cmidrule(lr){5-7}
    {Axiom Set} &  {Accuracy} &  {Precision} &    {Recall} &  {Accuracy} &  {Precision} &    {Recall} \\
    \midrule
         FLD1 &     0.885 &      0.857 &   0.923 &     0.947 &      0.918 &   0.982 \\
         FLD2 &     0.891 &      0.858 &   0.937 &     0.954 &      0.936 &   0.974 \\
         GEO6 &     0.829 &      0.792 &   0.893 &     0.850 &      0.822 &   0.893 \\
         GEO7 &     0.847 &      0.839 &   0.860 &     0.914 &      0.903 &   0.927 \\
         GEO8 &     0.812 &      0.776 &   0.876 &     0.880 &      0.842 &   0.934 \\
         GEO9 &     0.864 &      0.860 &   0.869 &     0.940 &      0.931 &   0.951 \\
         GRP5 &     0.763 &      0.747 &   0.796 &     0.860 &      0.864 &   0.856 \\
         KRS1 &     0.852 &      0.831 &   0.883 &     0.938 &      0.920 &   0.960 \\
         NUM9 &     0.848 &      0.829 &   0.876 &     0.836 &      0.812 &   0.875 \\
         SET1 &     0.793 &      0.804 &   0.774 &     0.818 &      0.805 &   0.837 \\
    \bottomrule
    \end{tabular}
\end{table}

\begin{table}[t!]
    \centering
    \caption{\label{table:success-counts}Comparison of the basic saturation prover, MLP-on-FwdP, GNN-on-FwdP and E. The test sets consisted of 606 human-generated theorems in total. E --auto is shown separately as a reference as it uses advanced heuristics unlike the other provers. Bold numbers show the best performance for each dataset.}

    \robustify\bfseries
    \sisetup{
        table-align-text-pre     = false,
        table-align-text-post    = false,
        input-open-uncertainty   = ,
        input-close-uncertainty  = ,
        detect-weight            = true,
        detect-inline-weight     = math,
    }
    \begin{tabular}{
        c
        S[table-format=3.0,round-mode=places,round-precision=0] |
        *4{S[table-format=3.0,round-mode=places,round-precision=0]} |
        S[table-format=3.0,round-mode=places,round-precision=0]
    }
    \toprule
    {Axiom Set} &  {Theorems} & {Basic} &  {MLP-on-FwdP} & {GNN-on-FwdP} &            {E} &  {E --auto} \\
    \midrule
           FLD1 &          78 &            9 &             20 &            28 &   \bfseries 29 &          45 \\
           FLD2 &         105 &            6 &             23 &            19 &   \bfseries 55 &          80 \\
           GEO6 &         128 &           78 &   \bfseries 83 &            69 &             79 &         128 \\
           GEO7 &          38 &           32 &             34 &  \bfseries 36 &             33 &          38 \\
           GEO8 &         128 &           78 &  \bfseries 100 &            80 &             79 &         128 \\
           GEO9 &          37 &           34 &   \bfseries 35 &  \bfseries 35 &             28 &          37 \\
           GRP5 &          10 &            4 &   \bfseries 10 &             7 &              5 &          10 \\
           KRS1 &          41 &            4 &             11 &  \bfseries 12 &   \bfseries 12 &          38 \\
           NUM9 &          30 &  \bfseries 4 &    \bfseries 4 &   \bfseries 4 &    \bfseries 4 &          14 \\
           SET1 &          11 &  \bfseries 6 &    \bfseries 6 &   \bfseries 6 &              5 &          11 \\
    \midrule
          Total &         606 &          255 &            326 &           296 &  \bfseries 329 &         529 \\
    \bottomrule
    \end{tabular}
\end{table}

\paragraph{Transfer from FwdP leads to better results than learning on TPTP only.} As a measure of transfer, we used the number of human-generated theorems proven by the learned models \emph{on top of what could already be proven by the basic saturation prover} (Table~\ref{table:confusion}). Overall, the MLP models trained on purely synthetic theorems (MLP-on-FwdP) were able to prove 32 more theorems than the MLP models trained on  human-generated theorems (MLP-on-TPTP). %
Similarly, GNN-on-FwdP were able to outperform GNN-on-TPTP on 5 axiom sets, proving 8 more theorems overall.

\begin{table}[t!]
    \centering
    \caption{\label{table:confusion}Comparison of the models trained on human-generated theorems (on-TPTP) and ones trained on synthetic theorems (on-FwdP). The test sets consisted of 351 human-generated theorems that could not already be proven by the basic saturation prover.}
    \robustify\bfseries
    \sisetup{detect-weight=true,detect-inline-weight=math}
    \begin{tabular}{
        c
        S[table-format=3.0,round-mode=places,round-precision=0] |
        *2{S[table-format=2.0,round-mode=places,round-precision=0]} |
        *2{S[table-format=2.0,round-mode=places,round-precision=0]}
    }
    \toprule
    {Axiom Set} &  {Theorems} &     {MLP-on-TPTP} &     {MLP-on-FwdP} &     {GNN-on-TPTP} &     {GNN-on-FwdP} \\
    \midrule
           FLD1 &        69.0 &             4 &  \bfseries 12 &             3 &  \bfseries 19 \\
           FLD2 &        99.0 &  \bfseries 21 &            17 &             6 &  \bfseries 15 \\
           GEO6 &        50.0 &   \bfseries 9 &             8 &  \bfseries 20 &             0 \\
           GEO7 &         6.0 &             2 &             2 &             1 &   \bfseries 4 \\
           GEO8 &        50.0 &             5 &  \bfseries 22 &  \bfseries 10 &             2 \\
           GEO9 &         3.0 &             2 &             2 &   \bfseries 3 &             2 \\
           GRP5 &         6.0 &             1 &   \bfseries 6 &             1 &   \bfseries 3 \\
           KRS1 &        37.0 &             0 &   \bfseries 8 &             0 &   \bfseries 8 \\
           NUM9 &        26.0 &             0 &             0 &             0 &             0 \\
           SET1 &         5.0 &   \bfseries 1 &             0 &   \bfseries 1 &             0 \\
    \midrule
          Total &       351.0 &            45 &  \bfseries 77 &            45 &  \bfseries 53 \\
    \bottomrule
    \end{tabular}
\end{table}

\begin{figure}
    \centering
    \input{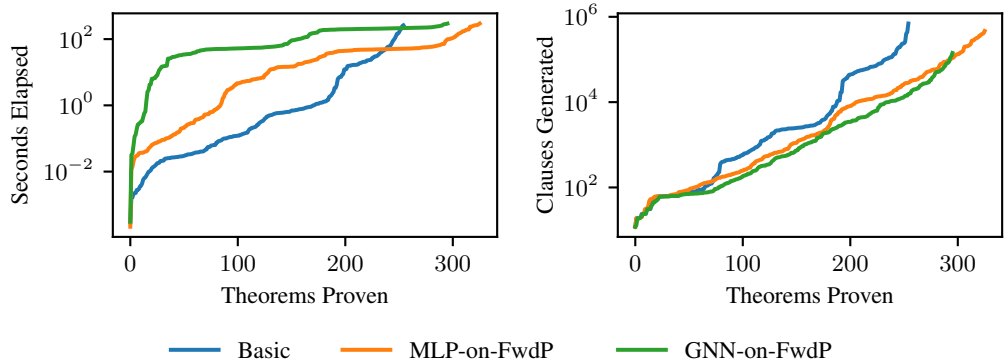}
    \caption{\label{fig:proof_stats} Search time and search steps for the basic saturation prover, MLP-on-FwdP and GNN-on-FwdP across all human-generated theorems. Provers with learned cost functions are significantly slower than the one based on clause weight only. Despite generating an order of magnitude fewer clauses during the search, these provers achieve better performance.}
\end{figure}

\paragraph{Learned provers are slower but stronger.} The improved results of the learned provers came despite the fact that they run slower than the basic saturation prover depending on the clause weight only. This was especially true for GNNs, which are on average 1633x slower than the basic prover and 10x slower than MLPs (Figure~\ref{fig:proof_stats}). MLPs generated an average of 21x fewer clauses than the basic prover. GNNs compensated for their slower wall-clock time by providing superior proof guidance and therefore generating fewer clauses in 8/10 datasets compared to MLPs.

\paragraph{Discussion.} Our results show that training a prover from naive proofs of synthetic theorems is feasible (Table~\ref{table:success-counts}). Additionally, training a prover on synthetic theorems can overcome the dataset size limitations posed by using only human-generated theorems (Table~\ref{table:confusion}). This is surprising, considering that the theorems generated by the forward proposer are not necessarily diverse nor similar to human-generated theorems. Also, relatively fewer clauses generated during a search for trained models (Figure~\ref{fig:proof_stats}) indicates that the models were able to distill and generalize information about clauses that appear in a proof. This way, the trained provers can save time by selecting clauses that are more likely to appear in proofs and therefore improve upon the weak prover which generated the training data. It should be noted that although we demonstrate that training a prover on synthetic data is very useful, we do not compete yet against the state-of-the-art heuristic provers like E. This work provides a new direction for building machine-learning-based theorem provers that can compete with heuristic provers in future.

\section{Related work}
\label{sec:related}

The most closely related work is ENIGMA~\citep{jakubuv&urban17} and the work of~\citet{loos&al17}; both learn a cost function for the saturation algorithm using machine learning. While ENIGMA uses a support vector machine (SVM) model, \citet{loos&al17} learn a neural network for classifying clauses. Our approach differs in two major ways. First, \citet{loos&al17} train the neural network on the human theorems and their proofs available in the Mizar dataset~\citep{kaliszyk&urban15} while we learn from synthetic data. Our approach is more effective in scenarios where only few proofs are available. Secondly, both these approaches \emph{embed} the learned cost function within the mature E theorem prover~\citep{schulz&al19}, thereby taking advantage of its many features and heuristics, whereas our approach uses minimal heuristic knowledge or extra features.

TRAIL~\citep{crouse&al19} is another recently proposed system that uses reinforcement learning for proof guidance. Their main goal is to build a domain agnostic prover which can generalize from Mizar to TPTP or vice versa. In contrast, our approach learns to transfer from synthetic theorems to TPTP and would be effective when no human theorems or proofs are available. Unlike all these works, we use the learned cost function with the vanilla saturation algorithm without any heuristics. 

\citet{wang&deng20} also propose a novel approach to prove theorems by generating theorems using reinforcement learning in Metamath~\citep{megill&wheeler19}. Their approach of learning a theorem generator requires an existing dataset of metamath proofs but results in more meaningful theorems that are similar to human-generated theorems. Their approach is useful only when a large dataset of human-generated theorems is available. In contrast, our method only uses axioms of a mathematical domain and can be effective even in the absence of any pre-existing theorems. We expect that our proposed technique can be used to augment the initial seed data for the method proposed in \citet{wang&deng20} for data scarce domains.

 In recent years, many systems have been proposed which illustrate the use of machine learning in automated theorem proving. A recent reinforcement learning based system, rlCoP~\citep{kaliszyk&al18},  uses Monte Carlo Tree Search (MCTS) along with neural networks in the (connection) tableau calculus \citep{hahnle01} (rather than the resolution calculus). The system was trained on human proofs in the Mizar dataset whereas our approach of using synthetic data is more effective when no proof data is available. FLoP~\citep{zombori&al19} optimizes towards being able to find longer proofs using reinforcement learning where evaluations are limited to synthetic theorems in Robinson's arithmetic. Several other recent works in higher order logic,
such as HolStep~\citep{kaliszyk&al17} and CoqGym~\citep{yang&deng19},
have attempted to create a new dataset of proofs and theorems or an environment like Gamepad~\citep{huang&al19} for machine learning but all these approaches rely on human generated proof data which is costly and may not be always available.

Finally, \citep{lample&charton20} use synthetic data in learning symbolic integration. Their approach outperforms popular symbolic integration systems like Mathematica on their own synthetically generated problems, but is not validated on an independent human-generated dataset.

\section{Conclusion}
\label{sec:conclusion}
This work proposes a novel approach of generating synthetic theorems by using forward inferences on axioms of any given domain of mathematics. Most of the previous work that uses machine learning in the ATP community depends directly on the existence of sufficiently large datasets of the human-generated theorems in order to train. We have shown that a synthetic dataset can be used instead and results in provers with good performance on human-generated data. 

We believe this novel approach opens up the possibility of moving the field of automated theorem proving from a data-scarce regime to a data-abundant regime. Particularly exciting are data augmentation techniques for automated theorem provers analogous to the computer vision domain. %
To accelerate the adoption of this technique, we will need to build smarter proposers. For example, our proposed dataset generates many more easy theorems than difficult ones, due to the random nature of the forward proposer. An approach that trains the proposer and the prover together in a reinforcement learning setting could allow us to exploit techniques from self-play
and curriculum learning.

\section*{Statement of broader-impact}

Our work is a step on the path towards improved proof automation through machine learning. While there is little doubt in many people's minds that machine learning is the future for automated theorem proving, today's state of the art systems are still built primarily on human prior knowledge rather than machine learning. Since we do not directly improve on the state of the art, there is little immediate practical impact of our work.

In the medium term, improved proof automation can help better verify hardware and software systems, giving us more reliable technology. Improved verifiability can help us avoid the considerable economic and personal losses that can result from malfunction and side-effects of these technologies. However, simply verifying that a property of a system holds does not guarantee that this property is the intended one, or that all unwanted behaviors have been ruled out. Instead, verification could lead to false confidence and provide an excuse to forgo the rigorous testing necessary to ensure the safety and reliability of our technology.

Theorem provers can also be used in planning and common sense reasoning to provide better routing information as well as answers to questions on complex datasets. We do acknowledge that such technologies, as with all technologies, can be misused for harmful intent. For example, datasets of personal information can be queried obtain information that can be used in a harmful way towards individuals or specific groups of people. 

In the long term, automated theorem proving has the potential to accelerate mathematical discovery and deepen our understanding of the world we live in.
Even problems in engineering or the sciences may be approachable, so long as they can be expressed in purely mathematical terms.
Moreover, a theorem prover doesn't need to be capable of solving difficult open problems on its own to be useful. Such a system can assist mathematicians in quickly exploring many simpler conjectures, and in formalizing existing theorems.

We believe that the social benefits in advances on many scientific fields enabled by advanced mathematical theorem provers as well as the benefits of more rigorous systems provide a \emph{net} positive effect on society.

\section*{Acknowledgements}
We would like to thank Natalie Lambert for helping us in planning our research goals and keeping our research focused during the project. We are also grateful to Koray Kavukcuoglu, Oriol Vinyals and Pushmeet Kohli for the helpful discussions during the course of this work. We would also like to thank Ramana Kumar, Adam Santoro, Arthur Guez and Rob Fergus for providing suggestions which helped in improving the manuscript substantially.

\bibliographystyle{named}
\bibliography{main.bib}
\appendix

\newpage
\setcounter{section}{0}
\section{Appendix: Details of the saturation algorithm}
\label{appendix:saturation}
In this section, we describe the details of our saturation algorithm. The pseudo-code can be found in Algorithm~\ref{alg:saturation}. The procedures $\theta$\code{-subsumption}, \code{find_resolutions}, and \code{find_factors} are the same as for other provers~\citep{riazanov&voronkov02,schulz&al19}. 

The main procedure is \code{saturation} and takes three inputs: the initial set of clauses (this includes axioms and negated conjecture clauses), a cost function\footnote{This can be a handcrafted heuristic or computed by a neural network.} which takes as input a clause and outputs its cost, and the age-cost ratio hyperparameter $a:c$. It maintains two priority queues at any given time: an age priority queue \code{qa} and a cost priority queue \code{qc}. The age priority queue is ordered solely by the iteration number at which a clause is generated, which ensures that every generated clause is processed after a finite number of iterations. The cost priority queue is ordered by the output of the cost function. The algorithm also maintains a set of processed clauses $P$. To begin with, all the initial clauses are inserted into both priority queues.

At each iteration of the algorithm, first we select a priority queue based on the age-cost ratio $a:c$: The age queue is selected for $a$ consecutive iterations, then the cost queue is selected for $c$ consecutive iterations an so on.
After selecting a queue, we select the clause $C_t$ that is at the top of this queue and remove it from both queues. If the clause is the empty clause, a refutation has been found and the theorem is proved.
Otherwise, the algorithm then conducts standard subsumption checks for the selected clause $C_t$ with the existing set of processed clauses $P$ (initially empty). Specifically, we check  forward and backward $\theta$-subsumption~\citep{plotkin1970note} to remove unnecessary clauses. Forward subsumption checks if the selected clause $C_t$ is less general than any clause in the processed set. 
A clause $C_1$ $\theta$-subsumes a clause $C_2$ if there exists a substitution $\theta$ that when
applied to $C_1$ gives $C_2$.
Circularity of subsumption checks is avoided by performing forward subsumption before backward subsumption.
To avoid another pitfall of subsumption checks where resolution can produce
clauses that increase in size but also in generality at the same time,
subsumed clauses are removed only if they pass an additional test: we say that a clause $C_1$ \emph{order-subsumes} a clause $C_2$
if $C_1$ has no more literals as $C_2$ and if $C_1$ $\theta$-subsumes $C_2$.
If the selected clause is subsumed by any existing clause in the processed set, 
it is simply discarded and we proceed to the next iteration with the appropriate age or cost queue. 
Otherwise, we proceed to check for backward subsumption: if any clause in the processed set $P$ is order-subsumed by $C_t$, it is removed from $P$.
Then, we compute all possible inferences (resolutions and factors) of $C_t$ with the remaining clauses in the processed set $P$, and the generated clauses are inserted in the age and cost priority queues. We also insert the clause $C_t$ in the processed set. The algorithm iterates until the queues are empty, which indicates that a refutation cannot be found and the initial set of clauses is satisfiable, meaning the theorem is not true. 

Observe that since all the initial clauses (axioms, hypotheses, etc.) are initially inserted in the priority queues, they are be subject to subsumption checks when selected. We also do simple syntactic tautology elimination. This is done by matching the negative literals of the clause to the positive literals of the clause syntactically. If there is a one to one match, the clause is marked as tautology and eliminated at the time of generation.

\newcommand{\activeset}{\text{active}}
\newcommand{\parents}{\text{par}}

\begin{algorithm}
\begin{lstlisting}
def order_subsumes($C_1$, $C_2$):
  return num_literals($C_1$) $\leq$ num_literals($C_2$) and $\theta$-subsumes($C_1$, $C_2$)

def saturation(initial_clauses, cost_fn, age_cost_ratio):
  a = numerator(age_cost_ratio)      # a in the ratio a:c
  c = denominator(age_cost_ratio)    # c in the ratio a:c
  qa = make_priority_queue(age)      # age queue of unprocessed clauses
  qc = make_priority_queue(cost_fn)  # cost queue of unprocessed clauses
  $P$ = {}                              # set of processed clauses
  qa.insert(initial_clauses)
  qc.insert(initial_clauses)
  t = 0
  while not qc.empty():
    if t %
      # Select the oldest unprocessed clause.
      $C_t$ = qa.extract_min()
      qc.remove($C_t$)
    else:
      # Select the unprocessed clause with the least cost.
      $C_t$ = qc.extract_min()
      qa.remove($C_t$)

    if is_empty_clause($C_t$):
      return "refutation_found"  # i.e. unsatisfiable

    # FORWARD SUBSUMPTION
    # Discard $C_t$ if it is order-subsumed by a clause in $P$.
    if $\exists C\in P$ s.t. order_subsumes($C$, $C_t$):
      continue

    # BACKWARD SUBSUMPTION
    # Discard any clause in $P$ that is order-subsumed by $C_t$.
    for $C$ in $P$:
      if order_subsumes($C_t$, $C$):
        $P$ = $P\setminus\{C\}$
  
    # Enqueue the factors of $C_t$ and any resolutions between $C_t$ and the clauses in $P$.
    new_clauses = find_factors($C_t$) $\cup$ find_resolutions($C_t$, $P$)
    qa.insert(new_clauses)
    qc.insert(new_clauses)

    # Add $C_t$ to the set of processed clauses.
    $P$ = $P\cup\{C_t\}$
    
    t = t + 1

  return "refutation_not_found"  # i.e. satisfiable
\end{lstlisting}
\caption{The saturation algorithm.}
\label{alg:saturation}
\end{algorithm}

\section{Appendix: Details on the Graph Representation}

\begin{figure}
    \centering
    \includegraphics[width=\textwidth]{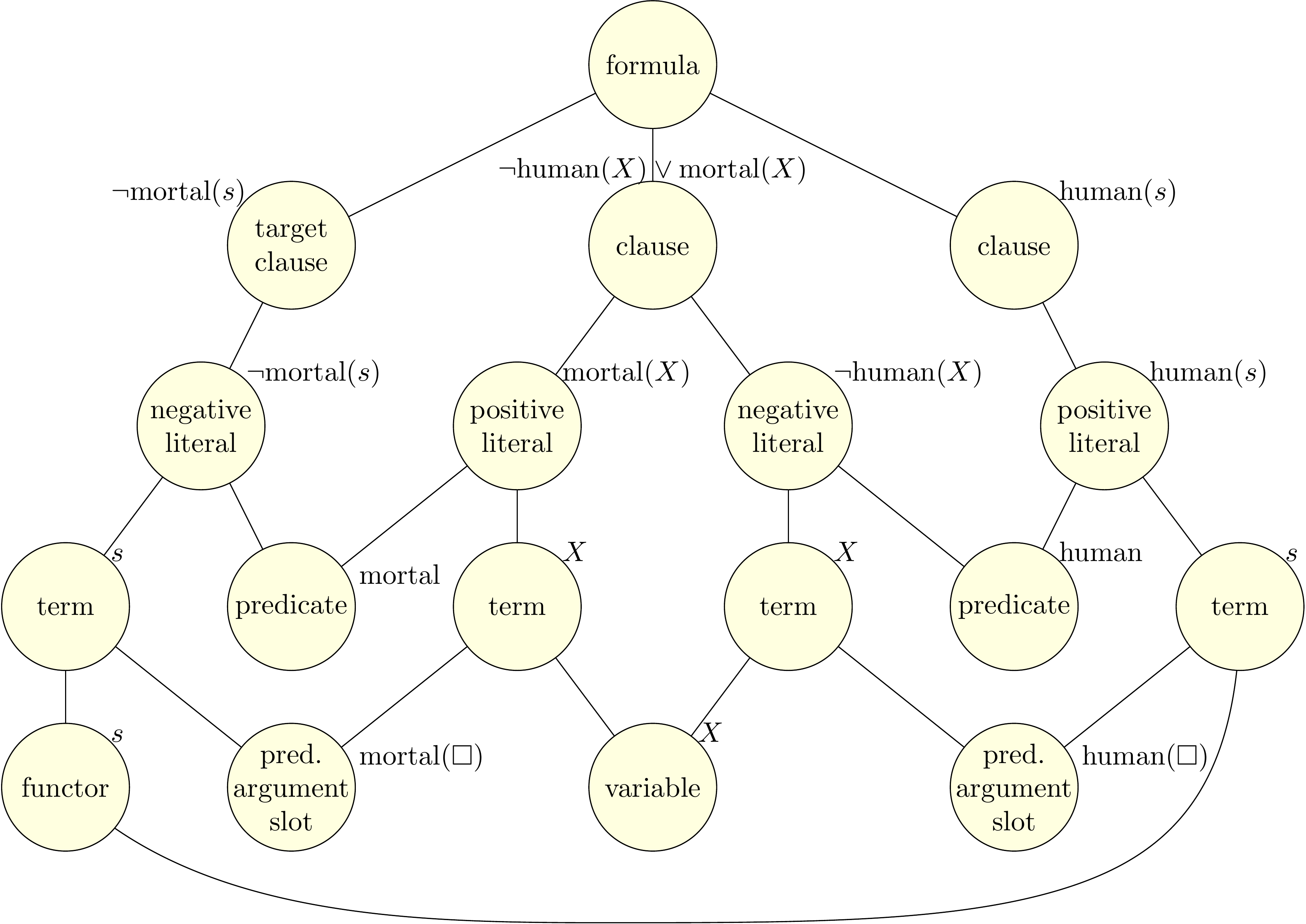}
    \caption{\label{fig:graph_repr} Graphical representation for the target clause $\neg \mathrm{mortal}(s)$ where the initial clauses are $\neg \mathrm{human}(X) \lor \mathrm{mortal}(X)$ and $\mathrm{human}(s)$. Text inside the nodes show the node types whereas text outside of the nodes show what each node represents. The graph is almost a tree except for variables, functors, predicates, and functor and predicate argument slots, which are shared by the nodes that refer to them.}
\end{figure}

In this section, we give an example of how a given clause in FOL can be converted into its graphical representation. The representation of the input where the initial clauses are $\neg \mathrm{human}(X) \lor \mathrm{mortal}(X)$ and $\mathrm{human}(s)$, and the target clause is $\neg \mathrm{mortal}(s)$ is shown in Figure~\ref{fig:graph_repr}. There are different node types in this graph representing different elements of a CNF formula such as formula, clause, positive and negative literals etc. The root node is a formula node which is the parent of all clause nodes. The target clause is represented by a special node type to distinguish it from the initial clauses. Each clause in turn is connected to two types of nodes: positive literals and negative literals. 

Each literal node is connected to a predicate node and one or more term nodes depending on the arity of the predicate. There will be a unique predicate node for each predicate and, hence, the predicate nodes are shared across literals and clauses. For example, the predicate node representing the predicate $\mathrm{human}$ is shared between literals of two clauses (Figure~\ref{fig:graph_repr}). Term nodes are one of the most complex types and can be connected to multiple types of nodes. First of all, each term node can be recursively connected to other term or variable nodes. Term nodes can also be connected to functor nodes. For example, the term node representing the term $s$ is connected to the functor node representing the functor $s$. A key difference between our representation and the representation of  \cite{glorotlearning} is the introduction of predicate and functor argument slot nodes. These nodes provide predicate (or functor) argument order invariance which we define as follows: If, in any FOL formula, the arguments of a predicate (functor) are permuted at all the occurrences of that predicate (functor), the formula remains invariant. To achieve this, we introduce argument slot nodes for each predicate and functor. The nodes for each argument slot of the same predicate or functor are unique in the whole formula; \emph{i.e.} these nodes are shared across clauses and literals. These argument slot nodes are also connected to term nodes and indicate which argument slot the current term occurs in its parent predicate or functor. In this example, we have predicate argument slot nodes for the only argument slots of both $\mathrm{mortal}$ and $\mathrm{human}$ connected to their corresponding terms. Constant (0-ary) functors do not have any slot nodes.

\section{Appendix: Additional Results}
We provide some more results in addition to results provided in Section \ref{sec:results}. We visualize and compare synthetic theorems and TPTP theorems on some simple statistics. We also provide the breakup of time taken and number of clauses generated per axiom set.

\subsection{Comparison of Synthetic theorems versus TPTP theorems}
\begin{figure}
    \centering
    \includegraphics{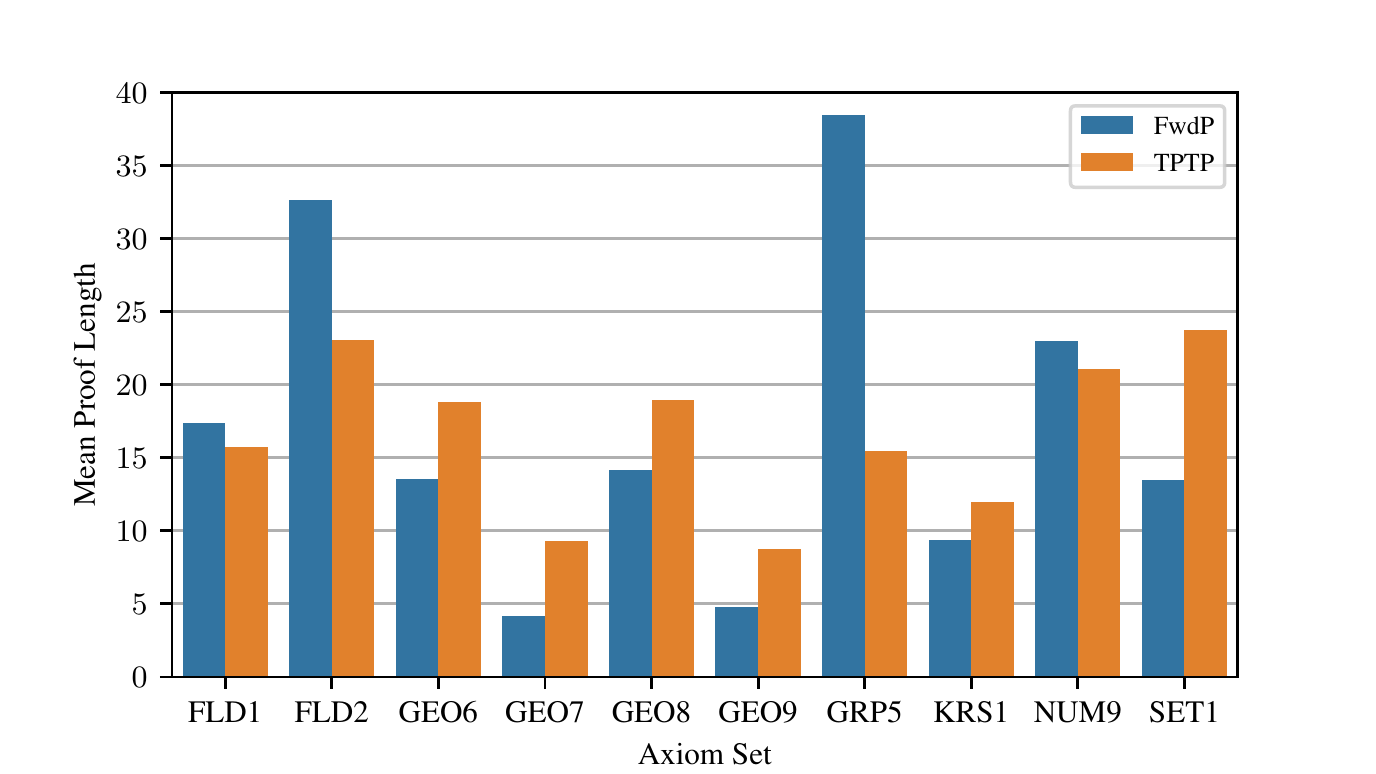}
    \caption{\label{appendixfig:dataset:proof_length} Comparison of mean proof lengths for proofs generated by E --auto for synthetic (FwdP) and human-generated (TPTP) theorems. For most of the axiom sets except GRP5, synthetic theorems are in a reasonable range compared to human-generated theorems.}

    \centering
    \includegraphics{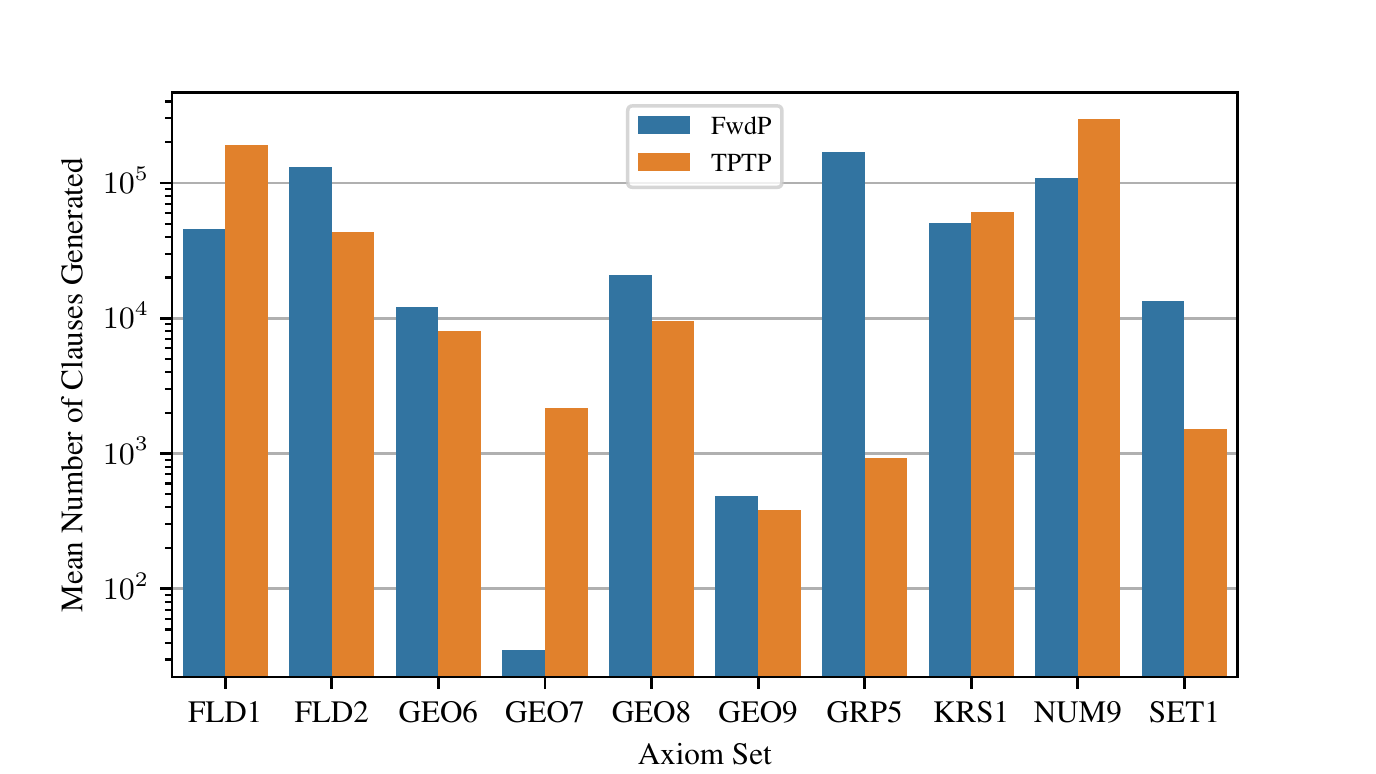}
    \caption{\label{appendixfig:dataset:num_clauses_generated} Comparison of mean of number of clauses generated by E --auto during the proof of synthetic (FwdP) and human-generated (TPTP) theorems. For many datasets, synthetic theorems require many more clauses than human-generated theorems}
\end{figure}

Although our synthetic theorems were generated without looking at TPTP theorems, in hindsight, we visualize how these synthetic theorems compare with TPTP theorems on some characteristics. %
We analyze the hardness or complexity of these two datasets by running E --auto on both the datasets. We extract the length of the proofs found by E for each theorem, and the number of clauses generated by E for each theorem. Although, this is not a true measure of complexity, we believe it is a reasonable proxy to understand the complexity or hardness characteristics of synthetic theorems. We plot mean proof length for E and mean number of clauses generated by E on all axiom sets respectively in Figure~\ref{appendixfig:dataset:proof_length} and  Figure~\ref{appendixfig:dataset:num_clauses_generated}. For proof length, synthetic theorems are most of the time reasonable (sometimes overestimate and sometimes underestimate slightly) compared to TPTP theorems except on GRP5 axiom set, where synthetic theorems have significantly longer proofs compared to TPTP theorems. The analysis on the number of clauses generated by E is much more different. Here, synthetic theorem are many times not in a reasonable range compared to TPTP theorems. It should be noted that both these statistics of proof length and number of clauses generated are not truly inherent characteristics of theorems but also depend on prover used. These statistics indicate that although our synthetic theorem distribution does not match the exact distribution of human generated theorems, the forward proposer theorems are in a reasonable range and definitely provide a good learning ground for the scenarios where no or limited human theorems are available.

\subsection{Average Running Time and Clauses Generated}
\begin{figure}
    \centering
    \includegraphics{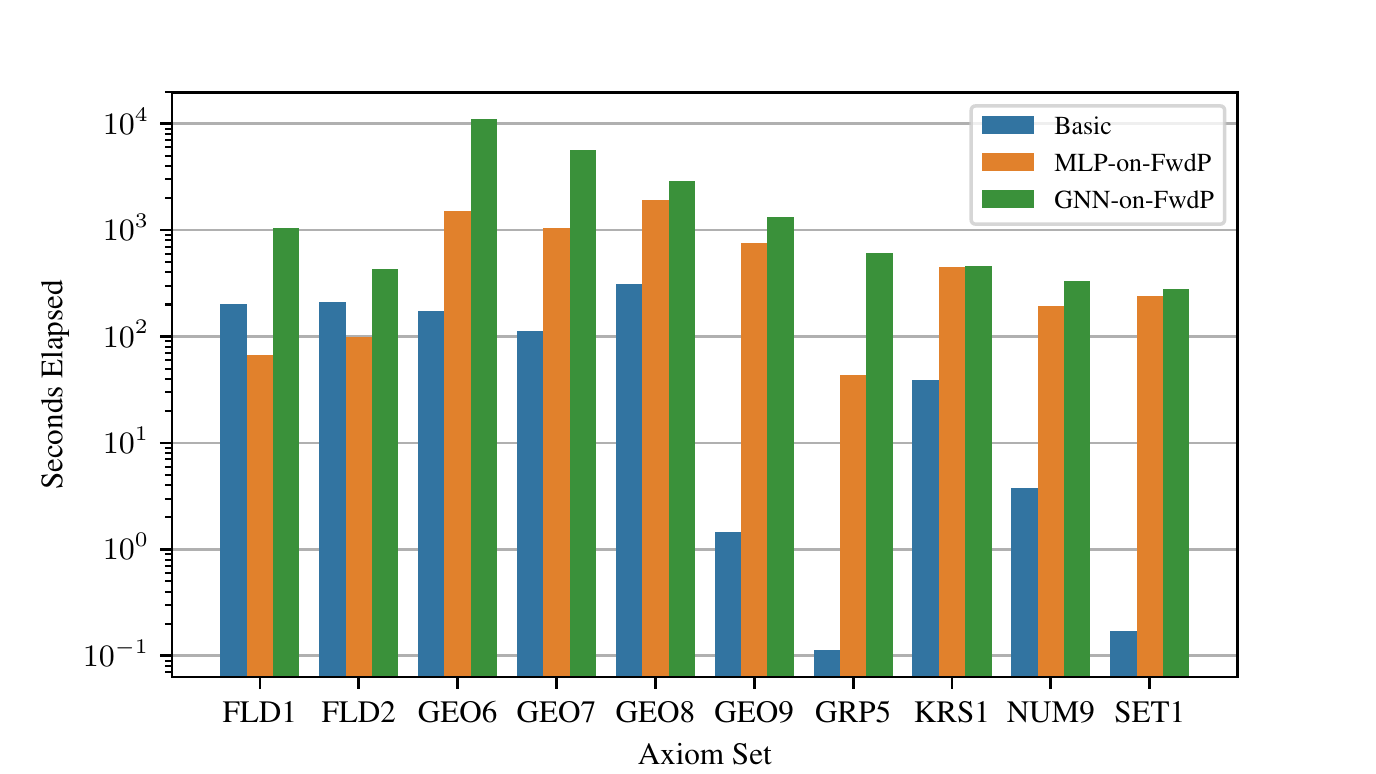}
    \caption{\label{appendixfig:seconds}Comparison of the total number of seconds elapsed during the proof of human-generated theorems (smaller is better). To make the numbers comparable, we used only the subset of theorems that can be solved by \emph{all} of the methods.}

    \centering
    \includegraphics{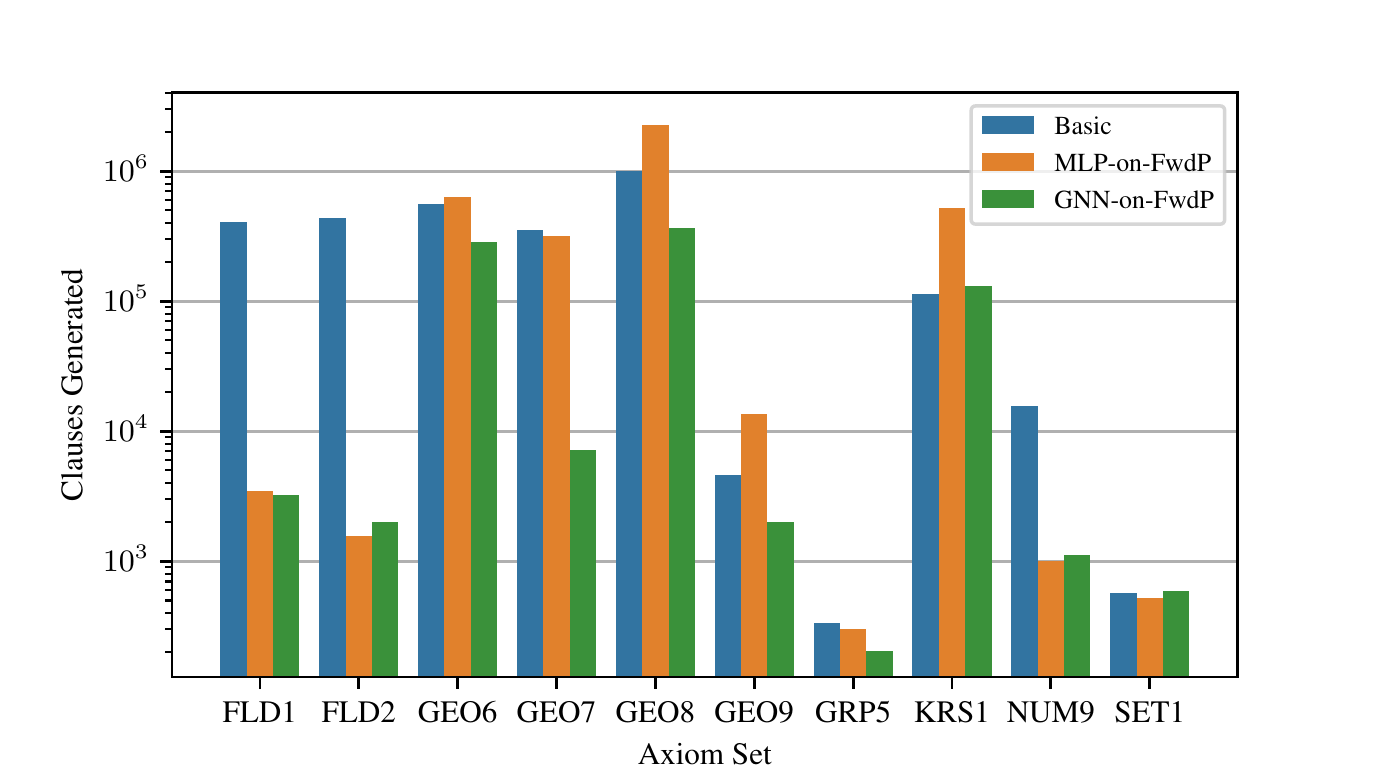}
    \caption{\label{appendixfig:clauses_generated}Comparison of the total number of clauses generated during the proof of human-generated theorems (smaller is better). To make the numbers comparable, we used only the subset of theorems that can be solved by \emph{all} of the methods.}
\end{figure}

Figure~\ref{fig:proof_stats} shows the overall time taken to solve problems and clauses generated for all problems in the axiom sets. This section breaks down the per-axiom results for time to solve (Figure~\ref{appendixfig:seconds}) and the number of clauses generated (Figure~\ref{appendixfig:clauses_generated}). In all axiom sets, learned cost functions took more time to solve problems than the heuristic. To compensate, the learned cost functions generated fewer clauses during the search for most datasets. Specifically, GNNs generate fewer clauses in 8/10 datasets.

\end{document}